\documentclass[12pt,preprint]{aastex}

\begin{document}
\title {A LOW SOLAR OXYGEN ABUNDANCE FROM THE 
FIRST OVERTONE OH LINES
}

\newbox\grsign \setbox\grsign=\hbox{$>$} \newdimen\grdimen \grdimen=\ht\grsign
\newbox\simlessbox \newbox\simgreatbox
\setbox\simgreatbox=\hbox{\raise.5ex\hbox{$>$}\llap
     {\lower.5ex\hbox{$\sim$}}}\ht1=\grdimen\dp1=0pt
\setbox\simlessbox=\hbox{\raise.5ex\hbox{$<$}\llap
     {\lower.5ex\hbox{$\sim$}}}\ht2=\grdimen\dp2=0pt\def\simgreat{\mathrel{\copy\simgreatbox}}
\def\simless{\mathrel{\copy\simlessbox}}
\newbox\simppropto
\setbox\simppropto=\hbox{\raise.5ex\hbox{$\sim$}\llap
     {\lower.5ex\hbox{$\propto$}}}\ht2=\grdimen\dp2=0pt
\def\simpropto{\mathrel{\copy\simppropto}}

\author{Jorge Mel\'endez\altaffilmark{1}}
\affil{Caltech, Department of Astronomy, M/C 105-24, 
1200 E. California Blvd, Pasadena, CA 91125}
\email{jorge@astro.caltech.edu}

\altaffiltext{1}{Affiliated with the Seminario Permanente de Astronom\'ia
y Ciencias Espaciales of the Universidad Nacional Mayor de San Marcos, Peru.}

\slugcomment{Send proofs to:  J. Melendez}

\begin{abstract}
An extremely high-resolution ($>$ 10$^5$) high-S/N ($>$ 10$^3$) solar spectrum 
has been used to measure 15 very weak first overtone ($\Delta v$ = 2)
infrared OH lines, resulting in a low solar abundance of $A_{O}$ $\approx$ 8.6 when
MARCS, 3D, and spatially and temporally averaged 3D model atmospheres are used. 
A higher abundance is obtained with Kurucz ($A_{O}$ $\approx$ 8.7) 
and Holweger \& M\"uller ($A_{O}$ $\approx$ 8.8) model atmospheres.
The low solar oxygen abundance obtained in this work 
is in good agreement with a recent 3D analysis of [\ion{O}{1}], \ion{O}{1},
OH fundamental ($\Delta v$ = 1) vibration-rotation and OH pure
rotation lines (Asplund et al. 2004). The present result brings further 
support for a low solar metallicity, and although using a low
solar abundance with OPAL opacities ruins the 
agreement between the calculated and the helioseismic measurement
of the depth of the solar convection zone, 
recent results from the OP project show that the opacities near the base 
of the solar convection zone are larger than previously thought, 
bringing further confidence for a low solar oxygen abundance.
\end{abstract}


\section{Introduction}

For many years, the accepted solar oxygen abundance
was $A_{O}$\footnote{$A_{X}$ $\equiv$ log ($N_X$/$N_H$) + 12} $\approx$ 8.9. This value was supported
by the impressive agreement between the
abundance obtained from the [\ion{O}{1}] lines (Lambert 1978, $A_O$ = 8.92 dex),
and the pure rotation (Sauval et al. 1984, $A_O$ = 8.91 dex)
and fundamental vibration-rotation OH lines (Grevesse et al. 1984, $A_O$ = 8.93 dex).
That solar oxygen abundance ($A_{O}$ = 8.9) is high
when compared to the abundance of nearby young OB stars
(e.g. Daflon et al. 2003, $A_O$ = 8.6-8.7 dex). 
Considering that the Sun is about 4.6 Gyr old, its
oxygen abundance is supposed to be lower or equal than the
present day interstellar medium (ISM). The ``high'' ($A_O$ = 8.9) solar 
oxygen abundance with respect to the ISM could be explained
by recent infall of metal-poor material into the local ISM (Meyer et al. 1994) 
or by migration (Wielen et al. 1996) of the Sun
from an inner (more metal-rich) birthplace in the Galaxy to its
present position.

Another alternative is to invoke errors in the oxygen abundance
derived in the Sun (or in OB stars).
The solar oxygen abundance has been recently revised (Allende Prieto et al. 2001;
Asplund et al. 2004; Allende Prieto et al. 2004;
hereafter AP01, A04 and AP04, respectively), 
yielding an abundance about 0.2-0.3 dex lower than previous determinations
(Sauval et al. 1984; Grevesse et al. 1984; Grevesse \& Sauval 1998).
Although this result solves the apparent oxygen
overabundance of the Sun with respect to the
local ISM,  it poses a new problem. The computed 
base of the solar convective zone (using the low value of 
the solar oxygen abundance) is then shallower than the
value obtained from helioseismology (Bahcall \& Pinsonneault 2004;
Bahcall et al. 2004a; Basu \& Antia 2004, Turck-Chize et al. 2004),
This new solar problem has been called by Bahcall et al. (2004b)
``the convective zone problem''. Furthermore, when low solar abundances
are employed, the computed sound speed radial profile disagrees
with sound speeds determined from helioseismological observations
(Bahcall et al. 2004a; Basu \& Antia 2004, Turck-Chize et al. 2004).
Therefore, it is important to verify the proposed low solar oxygen abundance
with other spectral features.

In the most comprehensive study utilizing different set of
lines available for oxygen abundance determination
in the Sun, namely the [\ion{O}{1}] (0.6 $\mu$m), 
\ion{O}{1} (0.6 - 0.9 $\mu$m), fundamental ($\Delta v$ = 1) 
vibration-rotation OH (3-4 $\mu$m), 
and pure rotation OH (9-13 $\mu$m) lines, 
A04 derived $A_{O}$ = 8.66 $\pm$ 0.05 dex
by using a 3D hydrodynamical model of the solar atmosphere.
The first overtone ($\Delta v$ = 2) OH lines (1.5-1.8 $\mu$m) were
probably not included in that work because they
are so weak that it is very difficult to identify and measure these lines 
in the solar spectrum\footnote{After receiving the first report from the
referee, we noted that first overtone OH lines in the
solar spectrum were first analyzed by Bonnell \& Bell (1993), in
a paper on ``The gravities of K giant stars determined from [\ion{O}{1}] 
and OH features''}. 
Note that in a recent 3D NLTE analysis of center-to-limb observations 
of the \ion{O}{1} infrared triplet (0.777 $\mu$m), AP04 have 
obtained $A_{O}$ = 8.70 $\pm$ 0.04 dex, this is somewhat
higher than the abundance obtained from permitted lines by
A04 ($A_{O}$ = 8.64). Considering the uncertainties
in the NLTE modeling, both results are equally plausible.

In this work we present measurements of extremely weak
first overtone infrared OH lines, which we
use for a new determination of the oxygen abundance in the Sun.

\section{Observational Data}

The solar spectrum we use was obtained with the 1-m FTS at the 
McMarth-Pierce telescope (Kitt Peak) by 
Wallace \& Livingston (2003, hereafter WL03). 
The center-disk solar
spectrum was obtained at very high spectral resolution 
(R = 700 000), and corrected for telluric lines (WL03). 
A portion of this spectrum is shown in Fig. 1.

The S/N of the spectrum is very high, it 
typically ranges from 900 to 1700, with a mean value of S/N = 1350,
as measured from 20 continuum regions in the $H$ band (1.5 - 1.8 $\mu$m).
This extremely low-noise spectrum allows the measurement
of features as weak as a few 0.1\% of the continuum.

The printed version of the solar atlas of WL03 provides 
identification for more than seventy OH lines of the $\Delta v$ = 2 
sequence, from the bands (2,0), (3,1), and (4,2).
The solar atlas of WL03 provides also identification
for some features that are blended with the infrared OH lines. 
For the less obvious blends, a line list of atomic and 
CN lines (Mel\'endez \& Barbuy 1999) was 
employed to avoid OH lines that are severely blended.
Also, a search for new OH features present in the solar
spectrum was undertaken using the list of OH lines by 
Goldman et al. (1998, hereafter G98). After careful examination, 
only 13 of the OH lines identified by WL03 were selected as 
suitable for an abundance analysis; the other lines
were discarded because of blending with known atomic or
CN lines (blends with unidentified absorption features
were also taken into account, considering only OH lines 
with  FWHM\footnote{Bonnell \& Bell (1993) reported an
average FWHM = 0.093 cm$^{-1}$ for 18 OH solar lines of the $\Delta v$ = 2 
sequence. This is equivalent to FWHM = 0.249 \AA\ , in excellent
agreement with our results (0.245 \AA)} 
= 0.245 $\pm$ 0.025 \AA).
Two additional infrared OH lines were identified in the
process. 

The equivalent widths of the OH lines are shown 
in Table 1, they were measured by fitting 
gaussian profiles with IRAF. 
The error in the measurement of these
weak features is about 10-20 \% (0.04-0.08 dex), 
as estimated from several trials using different continua. 
These errors are consistent with
an estimate obtained by using Cayrel's (1988) formula (7),
that results in $\delta W_\lambda$ = 0.11 m\AA\ ,
which for a typical 1-m\AA\ OH line corresponds to 
an error of 0.045 dex.

The errors in $W_\lambda$ could be also estimated
from a comparison with measurements by Bonnell \& Bell (1993), 
who used the infrared solar spectrum  of Delbouille \& Roland (1981), 
that has a lower spectral resolution by a factor of 2.33, but with
higher S/N by a factor of 2.96, therefore both spectra have
similar figure of merit $F \equiv$ (R $\times$ [S/N])/$\lambda$ (Norris et al. 2001),
although note that the Delbouille \& Roland solar atlas is not
corrected of telluric blends.
Bonnell \& Bell (1993) give $W_\lambda$ for 18 first-overtone OH lines 
in the $H$ band,
and eight of them are in common with our work. 
They estimated an uncertainty of 9\% (0.04 dex) for 
their measurements, but the standard deviation of the oxygen 
abundance obtained by them is 0.09 dex (23\%), 
therefore the errors of their $W_\lambda$ are probably similar to 
our measurements (10-20\%).
A comparison for the eight OH lines in common is shown in Fig. 2.
The mean difference (this work - theirs) is 0.07 m\AA\ ($\sigma$ = 0.19 m\AA), 
equivalent to a mean difference of 8\% ($\sigma$= 21\%), 
or 0.03 dex ($\sigma$ = 0.08 dex).


\section{Analysis}
The list of OH lines and their molecular parameters has been previously
described in Mel\'endez et al. (2001). It has been employed
to determine the oxygen abundance in halo (Mel\'endez et al. 2001)
and bulge stars (Mel\'endez et al. 2003). The
$gf$-values of the first overtone OH vibration-rotation 
lines ($X^2\Pi$) were computed from the Einstein $A$ coefficients
given by G98, that are the most
accurate available in the literature.
Moreover, the transition probabilities of G98 
are in good agreement with previous research (e.g.
Holtzclaw et al. 1993), although significant differences 
exist with respect to the earliest works (e.g. Mies 1974).
G98 and Holtzclaw et al. (1993) determined
Einstein $A$ coefficients employing the experimentally derived 
dipole moment function (DMF) by Nelson et al. (1990, hereafter N90).
N90 also determined $A$ coefficients, but
only for OH lines with low $J$ ($J"$ $\leq$ 14.5) numbers
(calculations for somewhat higher $J$ numbers were made available
by D. D. Nelson, Jr. 1999, private communication). 
Holtzclaw et al. (1993) noted that for lines of high $J$ (and high $v$),
an extrapolation of Nelson's DMF was necessary, although for the
lines used in the present work the differences due to this
extrapolation are negligible. In a different
approach, G98 used a theoretical DMF outside the
range of validity of Nelson's DMF, this allowed 
G98 to extend the calculation of Einstein coefficients 
to lines with very high $J$ and $v$. Note, however, that for the typical first
overtone OH lines observed in the solar spectrum (low $v$ and $J$), 
any of these approaches give essentially the same result.
G98 provides a comparison with the values
obtained by Holtzclaw and collaborators, and the agreement is
excellent. In Fig. 3c it is shown the difference between the 
log $gf$ value obtained by G98 and N90; 
the error bars were derived by N90, estimating the
95\% confidence limit in its experimental DMF,
taking into account both random and potential systematic errors. 
For the lines employed here, the error in the
$gf$-values is about 0.03 dex.

Line positions and excitation potentials are from G98,
they were calculated using the best available laboratory molecular
constants, and the line positions were checked with laboratory wavenumbers
(G98). The adopted value for the dissociation potential is 
4.392 eV (Huber \& Herzberg 1979). The parameters 
(wavelength, excitation potential, $gf$-value,
quantum numbers) of the set of lines used in this work are giving in Table 1.

The oxygen abundance was obtained using the 2002 version of the
program MOOG (Sneden 1973) and four model atmospheres: 
($i$) a Kurucz model\footnote{http://kurucz.harvard.edu}
including convective overshooting, although a
model with no overshooting (NOVER, Castelli \& Kurucz 2004) 
was also used, but the difference in the oxygen
abundance is only 0.01 dex  (lower with the NOVER model);
($ii$) the temporally and spatially averaged 3D solar model
\footnote{The use of this model was suggested by the referee} 
(hereafter $<$3D$>$ model) by A04;
($iii$) a MARCS solar model (Asplund et al. 1997),
($iv$) the Holweger \& M\"uller (1974, hereafter HM74) model atmosphere.
Additionally, M. Asplund (private communication, 2004)
have kindly made 3D calculations for a subset of the OH lines.

A microturbulence of 1 km s$^{-1}$ was adopted for the 1D modeling, 
yet the lines are so weak that the results do not depend on the
adopted microturbulence. The calculations were 
done in LTE; this seem to be a good approximation 
for vibration-rotation OH lines (Hinkle \& Lambert 1975).

The resulting abundances obtained with the $<$3D$>$ model 
are shown in Fig. 4 as a function of excitation 
potential, reduced equivalent width ($W_\lambda$/$\lambda$) and wavelength.
As can be seen, the derived oxygen 
abundances show no trends with either excitation potential
or wavelength. There is an apparent slight trend with reduced
equivalent width, but within the measurement errors ($\S$2)
this is not significant. The use of other model atmospheres
result in similar plots.

The mean value obtained from the 
15 OH lines and the $<$3D$>$ model is $A_{O}$ = 8.59 dex ($\sigma$ = 0.06 dex).
Full 3D calculations were kindly performed by M. Asplund for 5 lines
of the $\Delta$ v = 2 sequence with $\chi_{exc}$ = 0.30 - 1.24 eV.
After rescaling his results for the 15 OH lines, the full 3D result 
is 0.02 dex lower than the abundance obtained with the $<$3D$>$ model. 
The MARCS, Kurucz and HM74 models, give $A_{O}$ = 8.61, 8.71 and 8.80 dex, 
respectively. Our results are in good agreement (except the 
result with the HM74 model) with 3D hydrodynamical simulations of
other spectral features by AP01, A04 and AP04, who derived 
$A_{O}$ = 8.69 $\pm$ 0.05 dex, 8.66 $\pm$ 0.05 dex,
and 8.70 $\pm$ 0.04 dex, respectively. 

The main uncertainty in our results is the statistical error,
due to the weakness of the OH lines.
The only previous abundance analysis of the $\Delta$ v = 2 sequence
was performed by Bonnell \& Bell (1993). They analyzed
these lines in order to check the accuracy of the oscillator strengths, by
comparing the abundance obtained with these lines and the OH lines 
of the $\Delta$ v = 1 sequence. Bonnell \& Bell (1993) used the
HM74 model atmosphere and obtained an abundance of $A_{O}$ = 8.76 ($\sigma$ = 0.09 dex)
with the first overtone vibration-rotation OH lines, that
is 0.12 lower than the value they found with the fundamental OH lines.
After considering the mean difference in $gf$-values and equivalent
widths with respect to our work, their result is $A_{O}$ = 8.71 $\pm$ 0.09 dex,
which is lower than the abundance we found with
the HM74 model atmosphere ($A_{O}$ = 8.80 dex $\pm$ 0.06 dex). 
M. Asplund have also calculated the abundance of the first-overtone OH lines
using the HM74 model atmosphere and a subset of five lines, 
resulting in $A_{O}$ = 8.79 $\pm$ 0.06 dex (after normalizing his result
with the 15 OH lines). Considering systematic errors
of about 0.02 dex, and mainly considering the large statistical errors involved, 
the results with the HM74 model atmosphere are in reasonable agreement.

Before comparing our results with those obtained with other lines
analyzed with 3D hydrodynamical atmospheres,
we estimate the total error of our oxygen abundance for the 3D case.
The quadratic sum of the observational error ($\approx$ 0.06 dex),
the error of the $gf$-values ($\approx$ 0.03 dex), and a systematic
error of (at least) 0.02 dex, results in a total error of
$\delta A_{O}$ = 0.07 dex. 

\section{Comparison with 3D studies}
Considering that Kurucz model atmospheres 
are widely used in the literature, we have
also employed a Kurucz solar model to determine oxygen 
abundances using the [\ion{O}{1}], \ion{O}{1} and
$\Delta v$ = 0, 1 OH lines. For comparison purposes, the
abundance from these spectral features was also obtained 
using the $<$3D$>$ solar model. In short, the equivalent
widths and $gf$-values of the [\ion{O}{1}] and the \ion{O}{1} 777 nm triplet
was obtained from A04 (note that $W_\lambda$ for the
forbidden lines are cleaned from the \ion{Ni}{1} and CN blends),
and the equivalent widths for the $\Delta v$ = 1 and pure rotation
OH lines were measured from the WL03 and ATMOS solar atlas, respectively.
The $gf$-values for the molecular lines are from G98, which
are essentially identical to the $gf$-values employed by A04.
NLTE corrections for the \ion{O}{1} triplet were taken from A04.

In Table 2 a comparison of our results with those obtained
by A04 is shown. For the 3D case we can see that the molecular 
lines seem to give slightly lower abundances than the [\ion{O}{1}] and
\ion{O}{1} lines. This has been already noted by AP04, who
suggested that this difference could signal
that the results derived from the infrared OH lines
are less reliable than those obtained from the
forbidden and permitted lines.

Are these small differences in the abundance obtained
from different features due to errors in the transition
probabilities? The $gf$-values of the pure rotation OH lines
seem highly reliable.  In fact, an accurate dipole moment 
(1.6676 $\pm$ 0.0009 Debye) for $v$ = 0, $J$ = 9/2, 
have been measured by Meerts \& Dynamus (1973). As the vibration-rotation
interaction of the pure rotational lines is very small, an
extrapolation of the DMF (using theoretical calculations) 
for lines with higher $J$ and $v$ seems safe
(see e.g. Fig. 2 of Sauval et al. 1984). 
Goldman et al. (1983) adopted a
constant DMF for their analysis of the oxygen abundance
in the Sun from the pure rotation OH lines. They made a 
small correction to the experimental value derived by 
Meerts \& Dynamus (1973), to take into account the higher
$J$ values of the solar lines. In Fig. 3a we show the difference
between the $gf$-values derived by G98 and
Goldman et al. (1983), as can be seen, neglecting the variation
of the DMF results in an error of less than 1\%. 
A comparison between the $gf$-values obtained from G98
and Goorvitch et. al. (1992) is also shown is Fig. 3a.
Goorvitch et al. adopted Nelson's DMF, so the agreement with
G98 is not surprising.
The error bars shown in Fig. 3a. were estimated from the maximum error that is 
made if a constant DMF were adopted (for typical solar pure rotation 
OH lines with $v$ = 0, 1, 2, 3). An error of 0.01 dex for 
the $gf$-values of the $\Delta v$ = 0 sequence is not unreasonable.

The errors for the fundamental ($\Delta v$= 1) OH lines 
were estimated by N90,
as discussed in $\S$3. In Fig. 3b are shown these errors and the
difference between the $gf$-values derived from G98
and N90. In a different approach, G98 made
error estimates by making slight modifications to the shape of
the experimental DMF by N90. Using these results we obtained the
dashed error bars shown in Fig. 3b. The typical error for the
$gf$-values of the $\Delta v$ = 1 lines seem to be about 0.035 dex.

In $\S$3 we discussed the errors for the transition probabilities
of the $\Delta v$ = 2 OH lines, they are shown in Fig. 3c, and are
about 0.03 dex.

AP01 states that the uncertainty of the $gf$-values of
the forbidden lines is about 0.02 dex.
Lambert (1978) suggested log $gf$ = -9.75 for
the 6300 {\AA} line, 0.03 dex lower than the $gf$ value adopted by 
AP01 and A04, and the NIST database of critically selected 
transition probabilities suggests log $gf$ = -9.78, 0.03 lower
than the value recommended by Lambert (1978). We adopt
an uncertainty of 0.03 dex in the $gf$-value of the forbidden lines.
Another important uncertainty in the O abundance derived
from the 6300 and 6363 \AA\ [\ion{O}{1}] lines is due to blends 
with \ion{Ni}{1} and CN lines. In order to evaluate
these uncertanties, we have computed synthetic spectra of the CN
lines that are blended with the [\ion{O}{1}] lines. 
The contribution of CN lines is negligible for the 6300.3 \AA\
line, but it is very important for the 6363.7 \AA\ line. As already noted
by A04, the main blend is due to the (10,5) Q2 25.5 CN line.
Our calculation suggests a similar correction to that proposed
by A04 (0.5 m\AA). Fortunately, it is possible to observationally
check this result, by measuring the (10,5) Q1 25.5 CN line, which
obviously has the same excitation potential and $gf$-value as
the Q2 CN line. The Q1 CN line at 6365.0 \AA\ is apparently unblended, 
but the line is very weak and the continuum is difficult to define. 
After trying several solar atlases, we found that this line
was better defined in the Wallace et al. (1998) FTS solar spectrum
at disk center. When a correction to solar flux is made\footnote{A04
used the Kurucz et al. (1984) solar flux atlas, instead of a
disk center spectrum}, $W_\lambda$ = 0.35 m\AA\ is obtained\footnote{Note that if the 
true continuum is significantly higher than the local continuum, 
then this measurement is a lower limit}, which is
lower than the 0.5 m\AA\ employed by A04. The error in $A_O[6363]$ due to 
this uncertainty in the CN blend is 0.04 dex.
On the other hand, adopting a 20\% uncertainty for $W_\lambda$(\ion{Ni}{1}), 
results in an error of about 0.03 dex
in $A_{O}$ determined from the 6300 \AA\ line. These errors from 
the CN and \ion{Ni}{1} blends have been considered in column 5 of Table 3.

For the permitted lines, A04 used transition probabilities from the
NIST database. The difference between these oscillator strengths and
two different calculations given by Bi\'emont et al. (1991) is 0.025 dex,
so we adopt 0.03 dex as the error in the $gf$-values of the permitted lines.

The adopted errors are shown in Table 3, where are also shown the
statistical uncertainties ($\sigma$), the errors due to NLTE effects,
and estimates of other errors made in the modeling (see Table 3),
as well as the quadratic sum of the errors.
The error due to NLTE effects is only important for the \ion{O}{1} lines.
In particular, for the 777 nm triplet, the NLTE correction is
$\Delta \approx$ $-$0.21 dex, but it depends on the empirical
correction factor $S_H$. The NLTE correction could be
0.03 dex lower or higher, if $S_H$ = 0 (no collisions with H)
or $S_H$ = 1 (Drawin-like formula) is adopted, respectively (AP04).
Conservatively, we have adopted an NLTE error of 0.04 dex, 
slightly higher than this 0.03 dex uncertainty, and for
the abundance obtained from the permitted lines we
will consider the mean abundance obtained by A04 and AP04.

In Fig. 5 we show the oxygen abundances obtained 
from the different lines in the 3D analysis (Table 2),
and the error bars obtained above (Table 3). The recommended 
solar oxygen abundance by A04 ($A_{O}$ = 8.66 $\pm$ 0.05) is also
shown. As can be seen, within the errors, all different spectral
features, from the optical to the far infrared, 
are in good agreement. The weighted-average (using the
inverse square of the errors as weight) considering all the
lines is only 0.02 dex lower than the value recommended by A04.

\section{Summary and Conclusions}
A low oxygen abundance of $A_{O} \approx$ 8.6 dex (3D, $<$3D$>$ and MARCS models) 
was obtained from extremely weak (log ($W_\lambda$/$\lambda) \approx$ $-$7.2) 
first-overtone vibration-rotation OH lines present in the $H$ band.
These lines are about a factor of 5 weaker than
the fundamental vibration-rotation OH lines present in the $L$ band
of the solar spectrum. 
The OH lines used in this work are the weakest
ones ever used to obtain the oxygen abundance of the Sun
(e.g. for log ($W_\lambda$/$\lambda)$ = $-$7.2, an optical line
at 6000 \AA\ has $W_\lambda$ just 0.4 m\AA).

The present results support the low solar oxygen
abundance recently obtained from other spectral
features using 3D hydrodynamical model atmospheres
(as summarized in Table 2). 

Comparing the results obtained with different spectral features and
model atmospheres, we can see that the best agreement 
with the 3D simulations is obtained with the $<$3D$>$ model,
this is not unexpected, since this model represents
the temporal and spatial average of different
3D calculations. As stressed by A04, the $<$3D$>$ model cannot
reproduce all 3D effects, due to the lack of atmospheric
inhomogeneities; this can be seen comparing columns $<$3D$>$ and 3D
of Table 2. Note that the first overtone OH lines are the
less sensitive of the OH lines to the 3D effects, probably because
these lines are formed deeper than the fundamental and pure rotation
OH lines.

This new determination of the oxygen abundance in the Sun
is consistent with the abundance obtained in 
B and late O stars (Daflon et al. 2003), 
solving the problem of
the Sun being more oxygen-rich than the ISM. 
Nevertheless, the depth of the 
solar convection zone computed with the new solar
abundance and OPAL opacities (Bahcall \& Pinsonneault 2004) is 
in conflict with the measured value of the base of the convection
zone obtained by helioseismology (Basu \& Antia 1997, 
2004). Accordingly to Basu \& Antia (2004), the problem
could be solved with an upward revision of either 
the solar abundances or the opacity tables.
Bahcall et al. (2004a,b) have estimated that an increase of
about 10\% in the OPAL opacity tables is required to solve the
``convective zone problem'' and to bring the
computed sound speed profile into agreement 
with helioseismological observations.
Recently, Seaton \& Badnell (2004)
have shown that new calculations from the Opacity Project (OP)
result in opacities about 5\% larger than those given by OPAL
(for temperatures and densities near the base of the solar
convective zone), alleviating the difference
between solar models and helioseismology, and bringing support 
for a low solar oxygen abundance.

\acknowledgements
I thank the anonymous referee for his valuable suggestions
and noting recent pre-prints on helioseismology and OP opacities;
M. Asplund for his comments, sending the MARCS solar model and
performing 3D calculations; 
L. Wallace for providing a printed copy of the NSO solar atlas; 
and J. G. Cohen for her comments.
NSO/Kitt Peak FTS data used here were produced by NSF/NOAO.
I am grateful for partial support from NSF grant AST-0205951
to J. G. Cohen.

\clearpage

\begin{figure}
\epsscale{}
\plotone{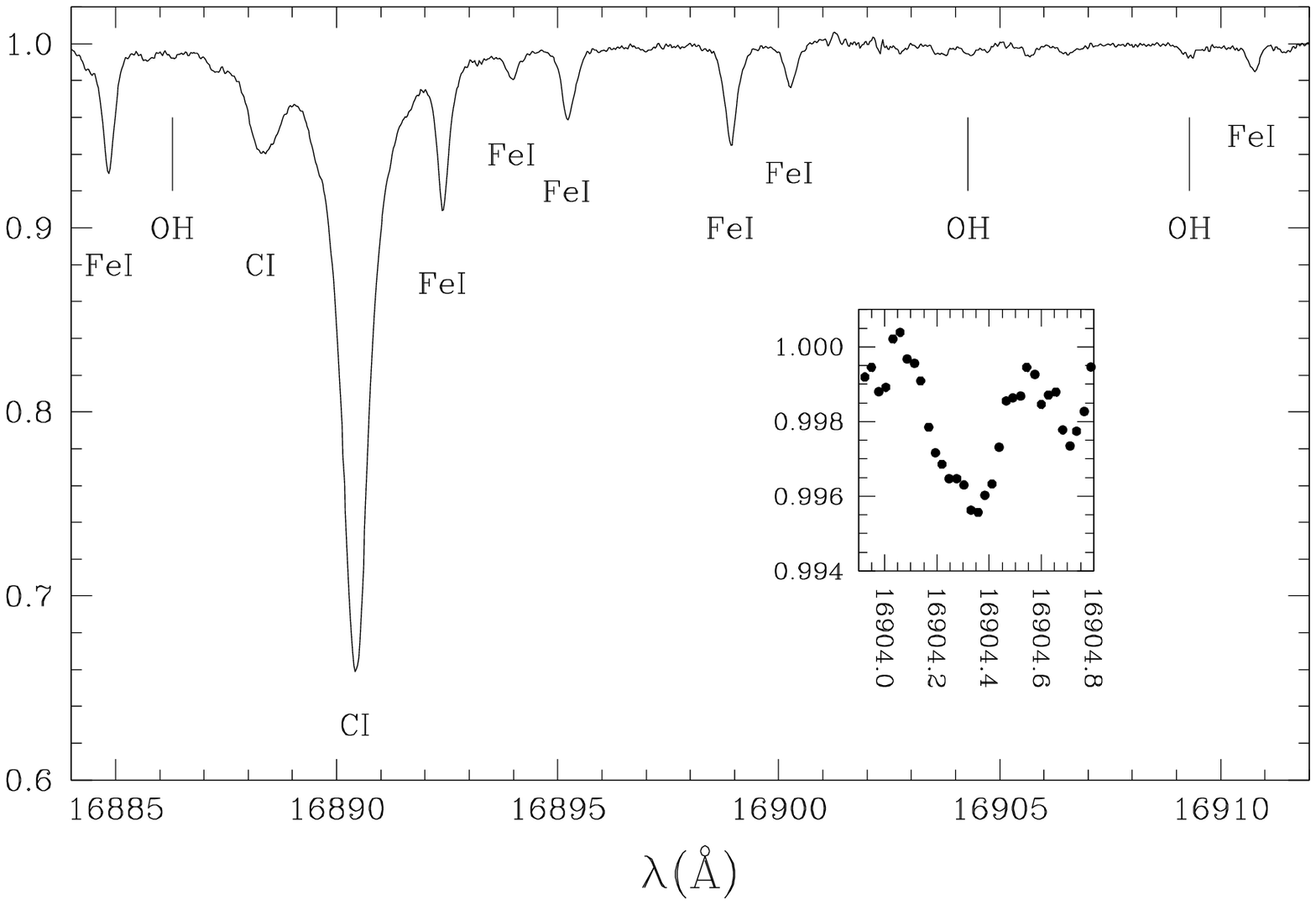}
\caption{Observed solar spectrum (WL03) around 1.69 $\mu$m. The vertical
lines indicate three OH lines used in this work. The strongest atomic 
features are identified. The profile of the OH line at 16904.3 \AA\ is 
shown in the inset, where only the upper 0.7 \% of the spectrum
is plotted.
}
\label{observed}
\end{figure}

\begin{figure}
\epsscale{}
\plotone{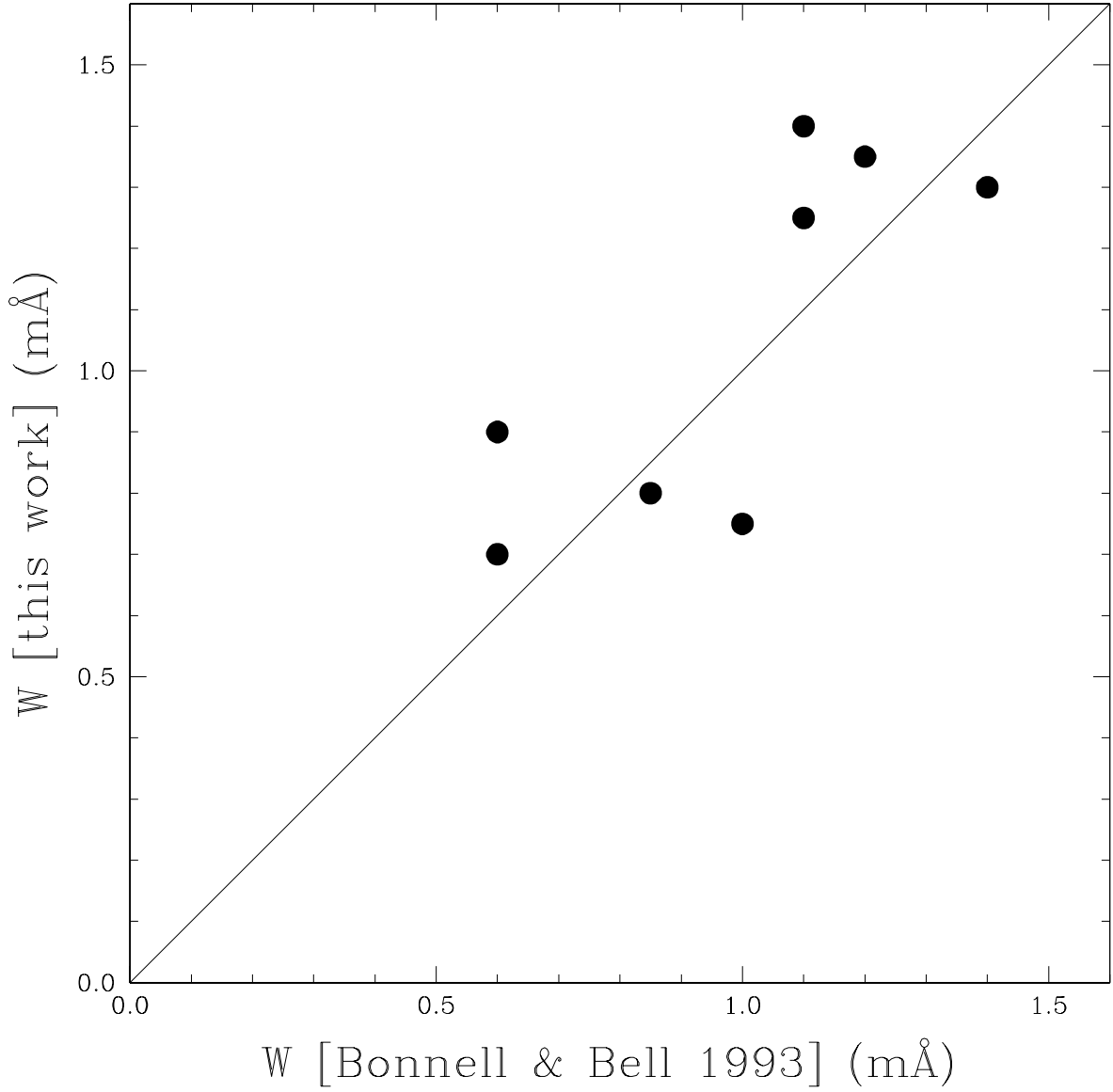}
\caption{Comparison between our $W_\lambda$ measurements and those by
Bonnell \& Bell (1993). The line depicts perfect agreement.
}
\label{ew}
\end{figure}

\begin{figure}
\epsscale{}
\plotone{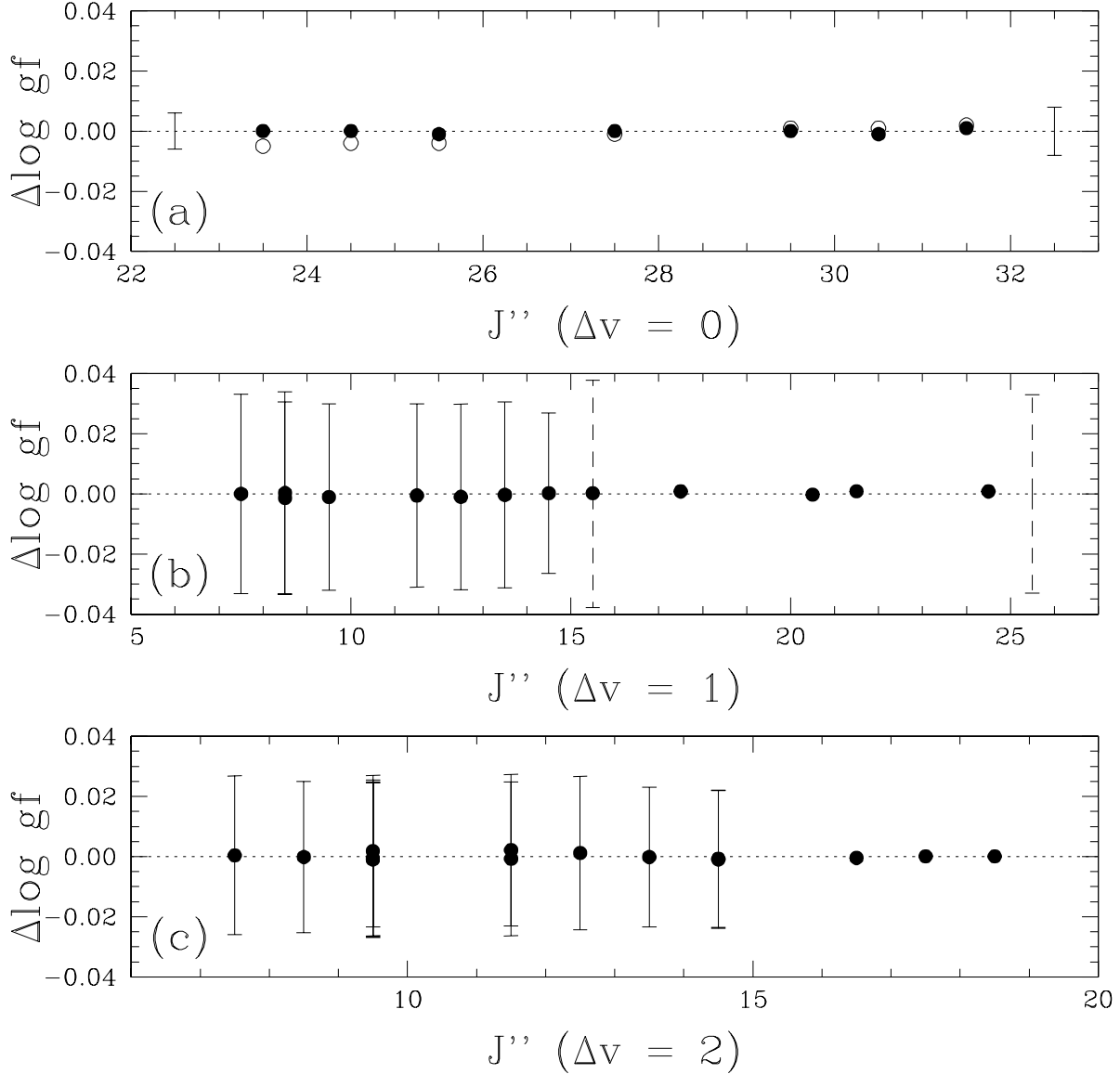}
\caption{Differences between the $gf$-values of infrared OH lines
calculated by G98 and 
(a) Goorvitch et al. (1992) (filled circles) and Goldman et al. (1983)
(open circles); (b) and (c) N90 (filled circles).
The error bars given in (a) were obtained supposing
that the DMF is constant. The solid error bars in (b) and (c) were estimated by
N90 taken into account the error of its experimental DMF.
The dashed error bars in (b) were estimated from slight variations to
the Nelson's DMF, as given in the last column of Table 3 by G98.
}
\label{gfs}
\end{figure}

\begin{figure}
\epsscale{}
\plotone{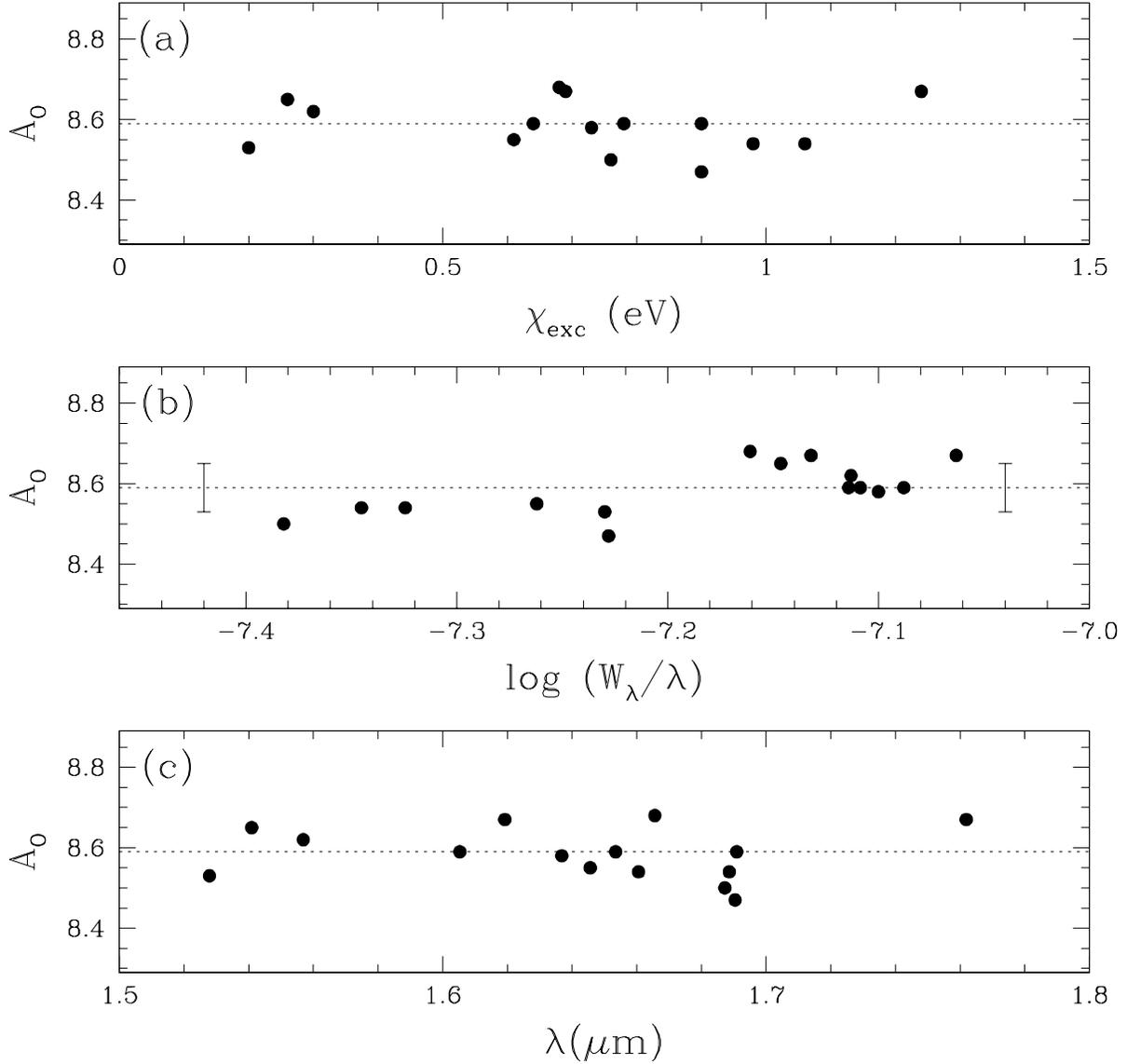}
\caption{Solar oxygen abundances obtained with the $<$3D$>$ model and 
first overtone infrared OH lines (filled circles) vs. (a) excitation potential,
(b) $W_\lambda$/$\lambda$, and (c) wavelength. The mean abundance ($A_{O}$ = 8.59 dex)
is shown by dotted lines, and the $\sigma$ (0.06 dex) error bar is shown
in the middle panel.
}
\label{newgf}
\end{figure}

\begin{figure}
\epsscale{}
\plotone{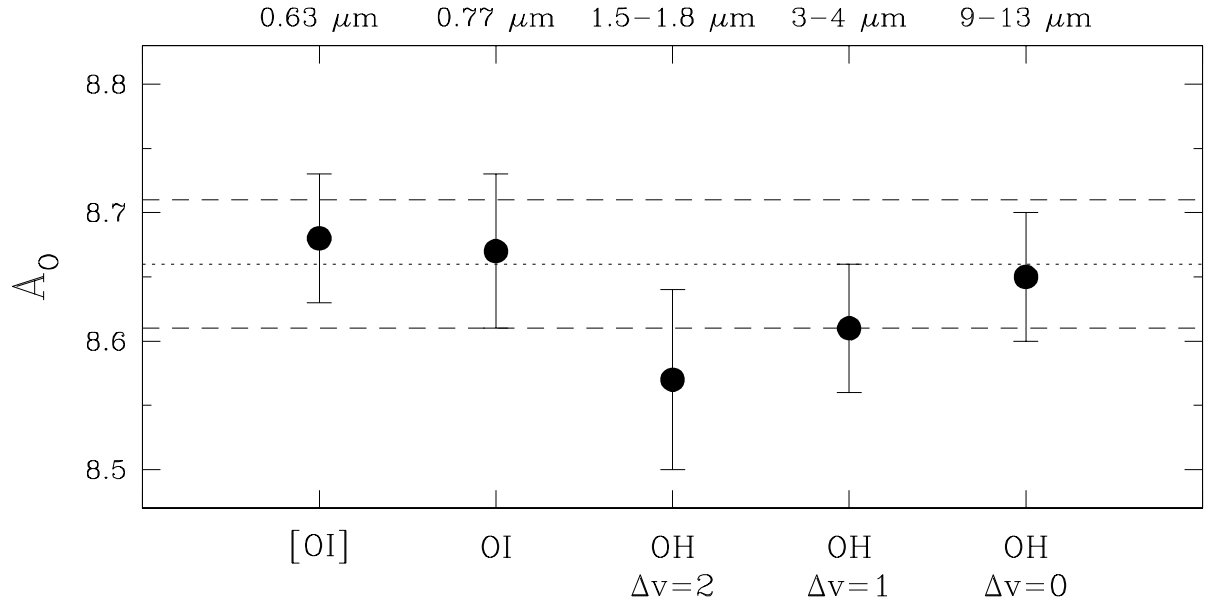}
\caption{Solar oxygen abundance derived from different lines in
3D analyses (Table 2). The error bars are from Table 3. The recommended 
value given by A04 ($A_{O}$ = 8.66 $\pm$ 0.05 dex) is shown by
the dotted and dashed lines.
}
\label{newgf}
\end{figure}

\begin{deluxetable}{llllrllllllr}
\tablecolumns{5}
\tablewidth{0pt}
\tabletypesize
\footnotesize
\tablecaption{Equivalent widths of OH lines}
\tablehead{
\colhead{$\lambda$\tablenotemark{a}}   &
\colhead{ID\tablenotemark{b}} &
\colhead{$\chi_{exc}$} &
\colhead{log {\it gf}} &
\colhead{$W_\lambda$} \\
\colhead{({\rm \AA})}   &
\colhead{}   &
\colhead{(eV)}  &
\colhead{(dex)} &
\colhead{(m{\rm \AA})} 
}
\startdata
  15278.52  &  (2,0) P1e 9.5  & 0.205  &  $-$5.382 &  0.9  \\
  15409.17  &  (2,0) P2e 9.5  & 0.255  &  $-$5.365 &  1.1  \\
  15568.78  &  (2,0) P1e 11.5  & 0.299  &  $-$5.269 &  1.2  \\ 
  16052.76  &  (3,1) P1e 9.5  & 0.639  &  $-$4.910 & 1.25  \\
  16192.13  &  (3,1) P2e 9.5  & 0.688  &  $-$4.893 &  1.4  \\
  16368.13  &  (3,1) P1f 11.5  & 0.731  &  $-$4.797 &  1.3  \\ 
  16456.04  &  (2,0) P1f 16.5  & 0.609  &  $-$5.048 &  0.9  \\
  16534.58  &  (3,1) P1e 12.5  & 0.782  &  $-$4.746 & 1.35  \\
  {16605.46}\tablenotemark{c} &  (4,2) P1e 7.5  & 0.982  &  $-$4.755 & 0.75  \\
  16655.99  &  (2,0) P1e 17.5  & 0.682  &  $-$5.010 & 1.15  \\
  {16872.28}\tablenotemark{c} &  (2,0) P1e 18.5  & 0.759  &  $-$4.975 &  0.7  \\ 
  16886.28  &  (4,2) P2e 8.5  & 1.059  &  $-$4.662 &  0.8  \\
  16904.28  &  (3,1) P1e 14.5  & 0.897  &  $-$4.654 &  1.0  \\
  16909.29  &  (3,1) P1f 14.5  & 0.898  &  $-$4.654 &  1.3  \\
  17618.89  &  (4,2) P1e 13.5  & 1.244  &  $-$4.403 &  1.3  \\
\enddata
\label{linelist}
\tablenotetext{a}{identifications in the solar spectrum by WL03}
\tablenotetext{b}{(v',v") branch J"}
\tablenotetext{c}{{lines identified in this work}}
\end{deluxetable}

\begin{deluxetable}{llllllll}
\tablecolumns{6}
\tablewidth{0pt}
\tabletypesize
\footnotesize
\tablecaption{Solar oxygen abundance from different spectral features and
model atmospheres.}
\tablehead{
\colhead{lines}   &
\colhead{3D}   &
\colhead{HM74} &
\colhead{MARCS} &
\colhead{$<$3D$>$} &
\colhead{Kurucz}
}
\startdata
{[\ion{O}{1}]}\tablenotemark{a}      & 8.68 $\pm$ 0.01 & 8.76 $\pm$ 0.02 & 8.72 $\pm$ 0.01 & 8.75 $\pm$ 0.01 & 8.79 $\pm$ 0.01 \\ 
 \ion{O}{1}\tablenotemark{a}          & 8.67\tablenotemark{c} $\pm$ 0.02 & 8.64 $\pm$ 0.08 & 8.72 $\pm$ 0.03 & 8.68 $\pm$ 0.03 & 8.67 $\pm$ 0.01 \\
OH ($\Delta v$ = 0)\tablenotemark{a} & 8.65 $\pm$ 0.02 & 8.82 $\pm$ 0.01 & 8.83 $\pm$ 0.03 & 8.74\tablenotemark{d} $\pm$ 0.05 & 8.80\tablenotemark{e} $\pm$ 0.06 \\
OH ($\Delta v$ = 1)\tablenotemark{a} & 8.61 $\pm$ 0.03 & 8.87 $\pm$ 0.03 & 8.74 $\pm$ 0.03 & 8.69 $\pm$ 0.06 & 8.82 $\pm$ 0.05 \\
OH ($\Delta v$ = 2)\tablenotemark{b} & 8.57 $\pm$ 0.06 & 8.80 $\pm$ 0.06 & 8.61 $\pm$ 0.06 & 8.59 $\pm$ 0.06 & 8.71 $\pm$ 0.06\\
\enddata
\tablecomments{The quoted uncertainties are only the $\sigma$ (line-to-line scatter)}
\tablenotetext{a}{Results from A04 for the 3D, HM74 and MARCS models, and
abundances obtained in this work for the $<$3D$>$ and Kurucz models}
\tablenotetext{b}{This work, except for the 3D case. The 3D calculations were performed
by M. Asplund, who obtained 8.58 $\pm$ 0.05 for 5 OH lines of the $\Delta v$ = 2 sequence,
using this result we found $A_{O}$ = 8.57 $\pm$ 0.06 for the 15 OH lines}
\tablenotetext{c}{This is the mean of the abundance obtained by A04 and AP04}
\tablenotetext{d}{$A_{O}$ = 8.75 $\pm$ 0.03 dex is obtained when only lines with $W_\lambda$ = 15-100 m\AA\ are employed}
\tablenotetext{e}{$A_{O}$ = 8.82 $\pm$ 0.03 dex for the lines with $W_\lambda$ = 15-100 m\AA}
\label{oxygen}
\end{deluxetable}

\begin{deluxetable}{llllllll}
\tablecolumns{6}
\tablewidth{0pt}
\tabletypesize
\footnotesize
\tablecaption{Errors in the oxygen abundance (dex)}
\tablehead{
\colhead{lines}   &
\colhead{$\sigma$\tablenotemark{a}} &
\colhead{$gf$s} &
\colhead{NLTE} &
\colhead{others\tablenotemark{b}} &
\colhead{total}
}
\startdata
{[\ion{O}{1}]}      & 0.01 & 0.03 & 0.00 & 0.04 & 0.05 \\ 
 \ion{O}{1}          & 0.02 & 0.03 & 0.04 & 0.02 & 0.06 \\
OH ($\Delta v$ = 0) & 0.01 & 0.01 & 0.00 & 0.05 & 0.05 \\
OH ($\Delta v$ = 1) & 0.03 & 0.04 & 0.00 & 0.02 & 0.05 \\
OH ($\Delta v$ = 2) & 0.06 & 0.03 & 0.00 & 0.02 & 0.07 \\
\enddata
\tablenotetext{a}{{line-to-line scatter}}
\tablenotetext{b}{minimum uncertainties in the 3D modeling, due to
systematic errors in the continuum opacities, partition functions, etc. 
For the [\ion{O}{1}] lines errors due to blends with \ion{Ni}{1}
and CN lines are considered.
A higher error for the pure rotation lines was adopted because
the 3D modeling is less successful for these lines (see Fig. 10 of A04)}
\label{oxygen}
\end{deluxetable}

\end{document}